# Nonreciprocal transport in a Rashba ferromagnet, delafossite PdCoO$_2$

*Jin Hong Lee,\* Takayuki Harada, Felix Trier, Lourdes Marcano, Florian Godel, Sergio Valencia, Atsushi Tsukazaki, and Manuel Bibes\**

KEYWORDS: *Rashba-type spin-orbit interaction, surface ferromagnetism, nonreciprocal transport, magnetoresistance, surface and interfacial phenomena, delafossite nanofilms*

ABSTRACT: Rashba interfaces yield efficient spin-charge interconversion and give rise to nonreciprocal transport phenomena. Here, we report magnetotransport experiments in few-nanometer-thick films of PdCoO$_2$, a delafossite oxide known to display a large Rashba splitting and surface ferromagnetism. By analyzing the angle dependence of the first- and second-harmonic longitudinal and transverse resistivities, we identify a Rashba-driven unidirectional magnetoresistance that competes with the anomalous Nernst effect below the Curie point. We estimate a Rashba coefficient of $0.75 \pm 0.3$ eV Å and argue that our results qualify delafossites as a new family of oxides for nano-spintronics and spin-orbitronics, beyond perovskite materials.

The broken space-inversion symmetry in condensed matter has enlarged the boundaries of quantum materials[1,2] by playing an important role in magnetoelectrics, topological electronics, spintronics and spin-orbitronics. Among a wide variety of phenomena related with inversion symmetry breaking, nonreciprocal (or directional) transport in noncentrosymmetric materials[3–6] has been explored in metals, semiconductors, topological insulators and superconductors. In particular, conductive surfaces and heterointerfaces have drawn attention due not only to their intrinsic/ubiquitous polar nature along the out-of-plane direction but also to the possibility to find high-mobility electrons by virtue of two-dimensional (2D) transport. In this regard, 2D electron systems, $e.g.$, topological insulator heterostructures[7] and oxide heterointerfaces[8–11], have been investigated in terms of unidirectional magnetoresistance (UMR)[12] originating mainly from the Rashba-type spin-orbit interaction (SOI). In such 2D Rashba systems, the spin and momentum of electrons at the Fermi level are nearly perpendicular to each other (spin-momentum locking[13]) and when an external in-plane magnetic field is applied, two resistive states are found in the magnetoresistance (MR) geometry[11,14]. This strong correlation between spin and momentum in the Rashba systems has been an essential element for spin-to-charge interconversion physics via the direct/inverse Rashba-Edelstein effects.[15,16] While the Rashba-type SOI is a universal feature of surfaces and heterointerfaces, transition metal oxides also often display unexpected magnetic properties at their surfaces and interfaces[17]. However, the interesting interplay that may arise between the Rashba-type SOI and surface ferromagnetism in a single compound, that would then qualify as a Rashba ferromagnet[18], has not been investigated yet.

The delafossite oxide PdCoO$_2$ (PCO) displays quasi-2D electron transport[19] with an amazingly long electron mean free path (~20 μm)[20], emergent ferromagnetism at its Pd-terminated surface[21] and a strong Rashba-like spin splitting in the CoO$_2$-derived surface states[22]. Figure 1a shows the layered crystal structure of PCO[23] where metallic Pd[+] sheets are separated by insulating $[CoO_2]^-$ layers. The crystal structure of PCO is trigonal (space group: $R\bar{3}m$). In terms of the distances between the nearest oxygen ions in the *ab*-plane (2.83 Å for PCO and 2.75 Å on average for sapphire), the lattice mismatch is about −2.9%. Judging from the diffraction pattern of the HRTEM of PdCoO₂/sapphire[24], the crystal lattice of our PCO films is relaxed on sapphire substrates. Recent efforts on stabilizing high-quality PCO thin films by pulsed laser deposition (PLD)[25,26] or molecular beam epitaxy[27,28] have offered a valuable chance to approach various strain states and thicknesses of the material as well as to investigate the surface or interfacial properties in ultrathin (*i.e.*, few nm) films by increasing the surface-to-bulk ratio of the sample. Surface ferromagnetism in PCO films was previously detected through thickness-dependent anomalous Hall effect (AHE) measurements and X-ray magnetic circular dichroism.[29]

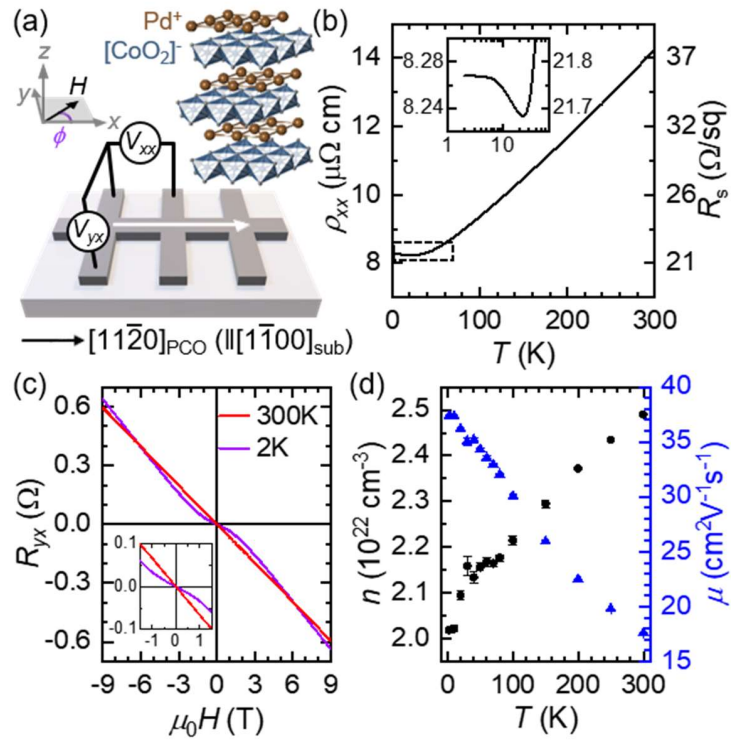

**Figure 1.** (a) Schematic diagram of PCO crystal structure and Hall-bar geometry for transport measurement. Basic characterization of the PCO device: (b) longitudinal resistivity, (c) Hall resistance, (d) carrier density and electron mobility as a function of temperature. Here, $\mu_0$ and $H$ denote the vacuum permeability and magnetic field strength, respectively.

In this Letter, we report magnetotransport experiments in an ultrathin PCO Hall-bar device as a function of external magnetic field, driving current and temperature in order to understand the origin of the nonreciprocal transport in the presence of both Rashba-type SOI and surface ferromagnetism. Below the magnetic transition temperature, we analyze the first- ($1\omega$) and second-harmonic ($2\omega$) components of $ac$ angle-dependent longitudinal MR and their scaling with current and magnetic field, to identify the UMR signal from other conventional MR components such as anisotropic magnetoresistance (AMR) and quadratic MR (QMR). By analyzing the sign and amplitude of observed UMR as a function of an external magnetic field at various temperatures, we discuss the coexistence of Rashba-type UMR[11] and anomalous-Nernst-(AN)-type electromotive force[24] where the former is a nonreciprocal transport mechanism and the latter is a thermal effect induced by unintentional Joule heating of the device. The observation of nonreciprocal responses in ultrathin high-mobility metals will provide a way to tackle pending questions on the delicate interplay between Rashba SOI and emergent phenomena at surfaces and also to design future spin-orbitronic or rectifying devices based on delafossite nanofilms.

Figure 1a shows a typical Hall-bar device for magnetotransport measurements (see MATERIALS AND METHODS for the details). As shown in Figure 1b,c, basic properties of electrical transport including longitudinal resistivity $\rho_{xx}$ (or sheet resistance $R_s$) and conventional Hall resistance $R_{yx}$ were obtained by using a four-terminal current-pulse measurement in the resistivity module of a Quantum Design Dynacool system. The resistivity curve shows a metallic behavior from room temperature to low temperature and a small upturn is observed at ~25 K (see the inset in Figure 1b). This upturn has been understood as a Kondo effect due to electron scattering from impurities or twin-domain boundaries of the PCO film[25] but its saturation behavior implies that the surface magnetic transition hinted from the s-shape of low-temperature Hall resistance curve (Figure 1c) may contribute as well. Figure 1d displays the charge carrier (electron) density $n$ calculated from the high-field (7 to 9 T) slope of the Hall resistance curves and the electron mobility $\mu$ estimated from the Drude model. The electron mobility of the sample ranges between 18 and 38 cm$^2$ V$^{-1}$ s$^{-1}$ from room to low temperature.

Bulk PCO is nonmagnetic as the B-site cation Co$^{3+}$ has a low-spin configuration. However, when it comes to ultrathin films, the type of surface termination strongly affects the magnetic properties of the system. Recent reports on the Pd-terminated polar surface of PCO have proven that a Stoner-like ferromagnetic state emerges at low temperature.[21,29] In order to visualize the magnetic transition at the surface of device, the AHE component $R_{yx}^{AH}$ as a function of temperature (Figure 2a) was obtained by subtracting the high-field slope from $R_{yx}$ curves as the ordinary term is evaluated at the high-field region owing to suppression of AHE component. Consequently, we can obtain the AHE signal, which will be proportional to the magnetization $M$ along the surface normal (or $z$ axis). The validity of this analysis was crosschecked by subtracting a reference $R_{yx}$ curve above the transition temperature $T_C$ from all other curves. Figure 2b summarizes the temperature-dependent amplitude of $R_{yx}^{AH}$ at 9 T. A steep change in the AHE amplitude appears at $T_C \sim 35$ K and $R_{yx}^{AH}$ saturates around 0.1 $\Omega$ at low temperature, which is consistent with the previously reported values of $R_{yx}^{AH}$ and $T_C$.[29]

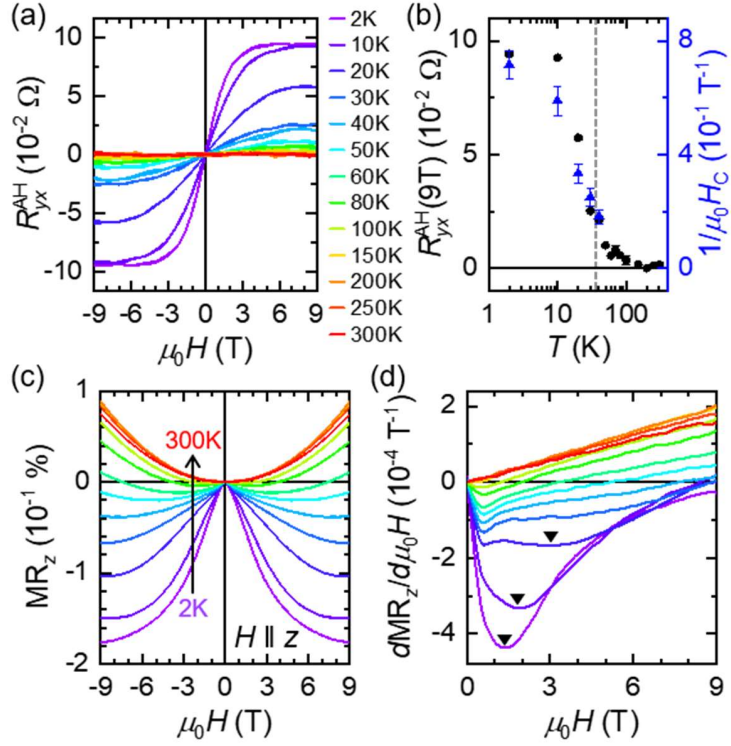

**Figure 2.** (a) Temperature-dependent AHE resistance. (b) AHE resistance extracted at 9 T representing surface ferromagnetic transition. A dashed line corresponds to the transition temperature. (c) Out-of-plane-field MR. Here, $MR_z = [R_{xx}(H) - R_{xx}(0)]/R_{xx}(0)$. (d) Derivative of $MR_z$ for the positive magnetic field. The reversed triangles denote the position of the broad minima ($\mu_0 H_C$) and they are summarized in (b).

Additionally, hints of surface ferromagnetism of the PCO device are also visible in the out-of-plane-field MR (or $MR_z$) measurement as shown in Figure 2c. Below $T_C$, we observe a nonlinear and negative MR[30], compatible with a surface magnetization with in-plane anisotropy, rotating to align with the external magnetic field. As the temperature increases, a parabolic dependence from Lorentz MR usually observed in metallic systems becomes dominant. The MR amplitude at 9 T is consistent with an electron mobility in the 30 cm² V⁻¹ s⁻¹ range. The derivative of $MR_z$ (or $d MR_z/d\mu_0 H$) plotted in Figure 2d shows two features; one is a sharp dip located at ~0.5 T meaning that the curvature of $MR_z$ curves changes from downward to upward, which seems to be related with the rotational motion of magnetization; the other is the broad minimum at $\mu_0 H_C$ (indicated by triangles) which we infer as an indicator of magnetic saturation. The trend of $1/\mu_0 H_C$ with temperature is overlaid in Figure 2b and matches well the surface magnetic transition deduced from AHE data.

We now turn to *ac* harmonic transport and to the detection of UMR to gain insight into the nonreciprocal transport response[12]. We apply an external magnetic field within the film plane and rotate it by an angle $\phi$ with respect to the direction of the *ac* current (see Figure 1a) and thereby break the time-reversal symmetry of the system. Depending on the angle between the external field and the spin polarization of flowing electrons, the amplitude of $V_{xx}$ varies and UMR is detected in the $2\omega$ channel as it depends on the direction of the driving current. On the other hand, the $1\omega$ channel detects reciprocal responses such as AMR or QMR[11]. Both $1\omega$ and $2\omega$ components of longitudinal MR were measured at 2 K with an *ac* current modulated by 77.00 Hz. Figure 3a shows the angle-dependent $1\omega$ component $\Delta R_{xx}^{1\omega}$ at magnetic fields of 0 to 9 T (an interval of 1 T) with an *ac* driving current of 1.4 mA. Figure 3b exhibits the same component at the *ac* current of 0.2-2.0 mA (an interval of 0.2 mA) with a magnetic field of 9 T. Figure 3c,d corresponds to the $2\omega$ component $\Delta R_{xx}^{2\omega}$ under the same condition. At first glance, both components have different periodicity in angle with a $\cos(2\phi)$ dependence for $1\omega$ signal and a $-\sin(\phi)$ dependence for $2\omega$ signal. Qualitatively, the difference comes from whether the signal has a linear or quadratic relation to the magnetization and/or magnetic field inside the system. From Figure 3e,f, we see that the amplitude of the $1\omega$ component $A_{xx}^{1\omega}$ (obtained by sine-function fitting of $\Delta R_{xx}^{1\omega}/R_0$ curves) is $H$-quadratic and current-independent, *i.e.*, consistent with AMR and/or QMR. On the other hand, the amplitude of the $2\omega$ signal $A_{xx}^{2\omega}$ behaves nonlinearly with the magnetic field and linearly with the current. This is different from the situation in conventional Rashba interfaces that display a linear scaling with both magnetic field and current (then the UMR is coined BMR, bilinear magnetoresistance[11]). A non-linear dependence with magnetic field on the nonreciprocal longitudinal resistance was previously observed at interfaces between non-magnetic and magnetic topological insulators[7], which also hints at a magnetic origin for our data on PCO films. It is possible to estimate the rectification coefficient $\gamma$ (see ref 5 for the definition) by using the slope of $A_{xx}^{2\omega}$ around $H = 0$ in Figure 3e, and our estimation shows $\gamma \sim 1.1 \times 10^{-3}$ A$^{-1}$ T$^{-1}$ comparable to the value obtained from a Bi helix[3].

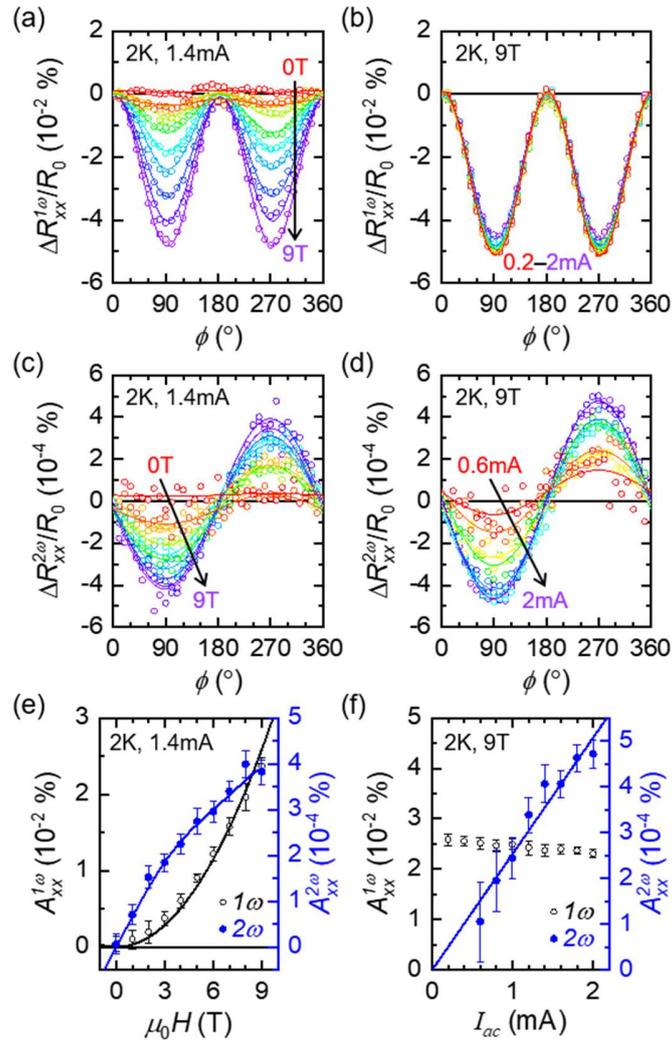

**Figure 3.** (a) Low-temperature (2 K) first-harmonic component of longitudinal MR as a function of angle at the magnetic field of 0 to 9 T. The amplitude of driving *ac* current is fixed at 1.4 mA.

Here, $\Delta R_{xx} = R_{xx}(\phi) - R_{xx}(0°)$ and $R_0 = R_{XX}^{1\omega}(0°)$. (b) Current dependence of the first-harmonic component at 9 T. (c-d) Second-harmonic component of longitudinal MR under the same measurement condition as (a-b). The circles and solid curves represent experimental data points and sine-function fits, respectively. Summary of magnetic-field (e) and driving-current (f) dependence of each component where $A_{xx}$ denotes the amplitude of the sine-function fits. The black curve in (e) corresponds to a quadratic fit to the first harmonic component, and the blue curve for the second harmonic component is obtained by sweeping the magnetic field while the angle is fixed. The solid line in (f) is the linear fit to the second harmonic component.

In order to clarify the origin of the observed UMR in the PCO device, the $2\omega$ component was investigated as a function of temperature across $T_C$. Surprisingly, the angle-dependent UMR data at various temperatures in Figure 4a show a sign reversal of the sinusoidal function at $T_C$. In order to avoid possible artifacts in the measurement geometry, the amplitude of the $2\omega$ signal was also monitored by sweeping the external magnetic field between ±9 T at a fixed angle, *i.e.*, 270° (Figure 4b). The field-swept $2\omega$ amplitude clearly shows that the low-temperature nonlinear curves not only reverse their sign but also become linear in shape, *i.e.*, corresponds to BMR, above $T_C$. The BMR then originates from the generation of a transverse magnetization $M_{DE}$ by the applied current due to the direct Edelstein effect[8]. The s-shape of nonlinear curves below $T_C$ is suggestive of its intimate relation with surface magnetism. The sign change in the data across $T_C$ points to an origin different from that of BMR.

When a ferromagnetic state appears in the system, AN effect may also arise due to a vertical thermal gradient along the $z$-axis generated by unintentional Joule heating produced by the current running into the device (Figure 4c); here, the *ac* current density in our device corresponds to $3.7 \times 10^{10}$ A m$^{-2}$ when we flow an *ac* driving current of 1.4 mA. It has been proposed that the comparison between the aspect ratio of the Hall bar $Z_H$ and the longitudinal-to-transverse ratio of the $2\omega$ component amplitudes $A_{xx}^{2\omega}/A_{yx}^{2\omega}$ allows an estimation of thermal effects[12,24]. In our case, $Z_H$ is 5 and $A_{xx}^{2\omega}/A_{yx}^{2\omega}$ is 4 at 2 K so that in this interpretation approximately 20% with the opposite sign of the observed signal would correspond to Rashba-type UMR (Figure 4c) while the rest would originate from AN-effect-driven electromotive force. The coexistence of AN- and Rashba-type transport becomes more visible if the $2\omega$ amplitude at 9 T is plotted as a function of temperature (Figure 4d). From room temperature down to $T_C$ the Rashba-type UMR component (*i.e.* BMR) is relatively independent on temperature but the AN response abruptly increase below $T_C$. We note that a possible contribution from ordinary Nernst effect in our measurement geometry can be discarded as the expected sign[26] is opposite to the observed BMR at room temperature.

Within this picture and using the microscopic model for BMR and QMR in a Rashba system (see ref 7 and equation 14 therein), it is possible to estimate the Rashba coefficient $\alpha_R$ from the observed ratio between the BMR and QMR amplitudes using physical parameters known for PCO (*cf.* ref 14 and Table 2 therein) as shown in Figure 4e. In consideration of the interlayer distance between conductive $Pd^+$ sheets, *i.e.*, 0.59 nm, and the thickness of the film, it is reasonable to assume that about one-sixth of the total current flows through the top-surface $Pd^+$ layer. Then, the value of the Rashba coefficient $\alpha_R$ at the surface is estimated to be $0.75 \pm 0.3$ eV Å in average above 70 K, which is one order of magnitude larger than the maximum value in LaAlO$_3$/SrTiO$_3$ interfaces[11]. Interestingly, Mazzola et al.[18] identified a Rashba-like spin splitting on the order of 50 meV at Pd-terminated surfaces by using nonmagnetic density-functional theory calculation but its role on the surface ferromagnetism has been elusive in the angle-resolved photoemission spectroscopy (ARPES) data due to much larger energy scale (~430 meV) of exchange splitting due to the Stoner transition appearing simultaneously at the surface. If we assume the Rashba-like momentum splitting $\Delta k_F$ at the Pd-terminated surface is on the order of 0.1 Å$^{-1}$, similar to what has been reported in the $CoO_2$-terminated case[19], it leads to the Rashba coefficient of ~0.5 eV Å compatible with the value obtained from our transport measurements. Another possible scenario for the observed Rashba-type magnetotransport is a strong interaction between the itinerant surface Pd ferromagnetism and the sub-surface $CoO_2$ layer which may still show a large Rashba-like spin splitting; as ARPES is sensitive to the top-most layer[18], it has been difficult to detect a splitting in the sub-surface region. Whether one mechanism is more dominant or both scenarios cooperate at the surface goes beyond our measurement limit and remains interesting, our result exemplifies a ferromagnetic surface showing the Rashba-type UMR due to the space-inversion symmetry breaking of a nonmagnetic quasi-2D metal.

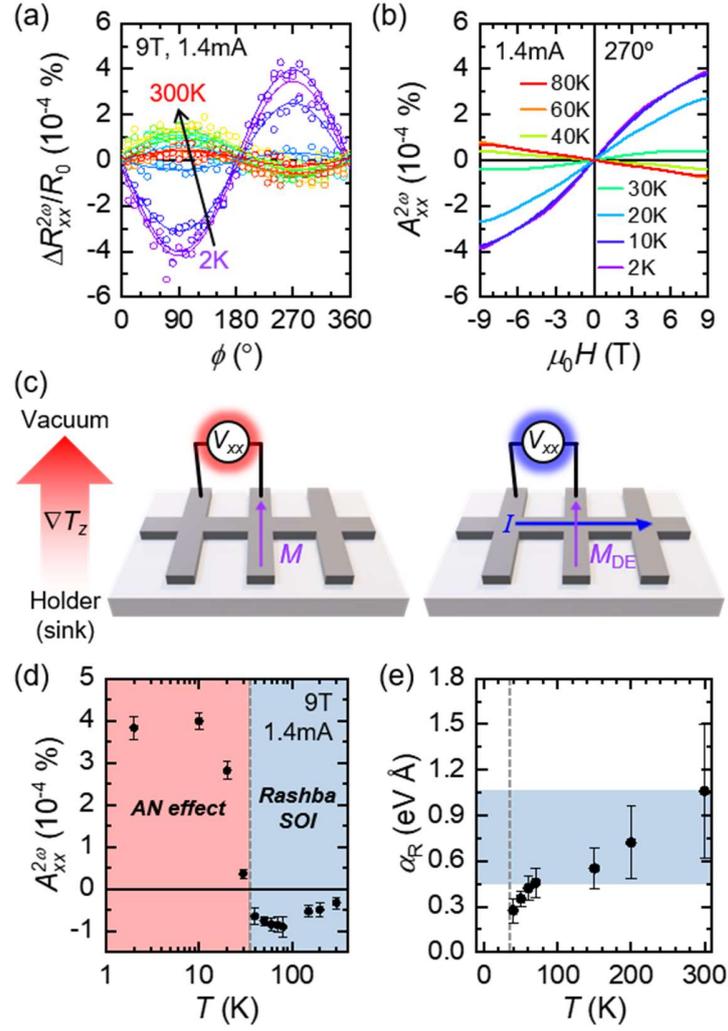

**Figure 4.** (a) Temperature-dependent second-harmonic component from 2 to 300 K where the color code is the same as that of Figure 2a. (b) Magnetic-field-swept second-harmonic component at a fixed angle. (c) Two possible origins of second-harmonic component: AN effect and Rashba SOI. (d) Summary of the second-harmonic amplitude as a function of temperature showing the dominant contribution at the surface. (e) Estimation of Rashba coefficient above surface magnetic transition $T_{\mathrm{C}}$.

Our study of nonreciprocal transport at the ferromagnetic surface of an ultrathin PCO film suggests that delafossite heterostructures will provide a fruitful chance to deepen our knowledge about Rashba-related phenomena, and related efforts on the stabilization of high-quality delafossite films have been progressing[31,32]. The unique coexistence of surface ferromagnetism and Rashba spin-orbit coupling offers exciting opportunities for spin-orbitronics and to test theoretical predictions such as skyrmionic spin-textures[18], signatures of which may have been observed from topological Hall effect in ref[29]. The delafossite family comprises several other interesting compounds, including good thermoelectrics and multiferroics such as $CuFeO_2$[33] or $AgCrO_2$[34]. Delafossites thus appear as a new family of quantum oxide materials from which complex multifunctional nanofilm heterostructures could be synthesized and lead to exotic nanoelectronic and spintronic phenomena, beyond the realm of oxide perovskites[35].

MATERIALS AND METHODS

A Hall-bar device (Figure 1a) was fabricated from a PLD-grown 3.8-nm-thick PCO thin film on a (0001)-oriented $Al_2O_3$ substrate (see ref 19 for the details of PLD growth) by utilizing optical lithography and $Ar^+$ ion-beam etching; here, there are six metallic $Pd^+$ sheets inside the film as the spacing of the sheets is 0.59 nm. The width of the Hall bar is 10 μm and the length between the longitudinal voltage ($V_{xx}$) leads is 50 μm. A $dc$ electrical current of 200 μA is applied parallel to the edge direction ($[11\bar{2}0]_{PCO}$) of triangular $Pd^+$ lattices.

AUTHOR INFORMATION

**Corresponding Author**

Jin Hong Lee – Unité Mixte de Physique, CNRS, Thales, Université Paris Sud, Université Paris-Saclay, F-91767 Palaiseau, France; Email: jinhong.lee@cnrs-thales.fr

Manuel Bibes – Unité Mixte de Physique, CNRS, Thales, Université Paris Sud, Université Paris-Saclay, F-91767 Palaiseau, France; Email: manuel.bibes@cnrs-thales.fr

**Authors**

Takayuki Harada – MANA, National Institute for Materials Science, 1-1 Namiki, Tsukuba, Ibaraki 305-0044, Japan

Felix Trier – Department of Energy Conservation and Storage, Technical University of Denmark, 2800 Kgs. Lyngby, Denmark

Lourdes Marcano – Dpto. Electricidad y Electrónica, Universidad del País Vasco-UPV/EHU, 48940 Leioa, Spain; Helmholtz-Zentrum Berlin für Materialien und Energie, Albert-Einstein-Straße 15, 12489 Berlin, Germany

Florian Godel – Unité Mixte de Physique, CNRS, Thales, Université Paris Sud, Université Paris-Saclay, F-91767 Palaiseau, France

Sergio Valencia – Helmholtz-Zentrum Berlin für Materialien und Energie, Albert-Einstein-Straße 15, 12489 Berlin, Germany

Atsushi Tsukazaki – Institute for Materials Research, Tohoku University, Sendai, Japan; Center for Spintronics Research Network, Tohoku University, Sendai, Japan

**Notes**

The authors declare no competing financial interest.

ACKNOWLEDGMENT

This work received support from the ERC Advanced grant n° 833973 "FRESCO". This work was partly supported by JST PRESTO (grant number JPMJPR20AD) and JST CREST (grant number JPMJCR18T2). F. Trier acknowledges support by research grant 37338 (SANSIT) from VILLUM FONDEN. L. Marcano acknowledges the financial support provided through a postdoctoral fellowship from the Basque Government (POS-2019-2-0017).